\documentstyle[12pt]{article}

\begin{document}

\hspace*{10cm} {\bf AMUP-94-09}

\hspace*{10cm} {\bf US-94-08}

\hspace*{10cm} {\bf December 1994}\\[.3in]

\begin{center}

{\large\bf Quark Mass Matrix with a Structure of a}\\[.1in]
{\large\bf Rank One Matrix Plus a Unit Matrix}
\footnote{
To be published in Mod. Phys. Lett. (1995).}
\\[2cm]

{\bf Hideo Fusaoka}\footnote{
E-mail: fusaoka@amugw.aichi-med-u.ac.jp} \\[.1in]

Department of Physics, Aichi Medical University \\
Nagakute, Aichi 480-11, Japan \\[.2in]

and \\[.2in]

{\bf Yoshio Koide}\footnote{
E-mail: koide@u-shizuoka-ken.ac.jp} \\[.1in]

Department of Physics, University of Shizuoka \\
52-1 Yada, Shizuoka 422, Japan \\[.4in]

{\large\bf Abstract}\\[.1in]

\end{center}

\begin{quotation}
A quark mass matrix model $M_q=M_e^{1/2} O_q M_e^{1/2} $ is proposed
where $M_e^{1/2}={\rm diag}(\sqrt{m_e},\sqrt{m_\mu},\sqrt{m_\tau})$ and
$O_q$ is a unit matrix plus a rank one matrix.
Up- and down-quark mass matrices $M_u$ and $M_d$ are described
in terms of charged lepton masses and additional three parameters
(one in $M_u$ and two in $M_d$). The model can predict reasonable
quark mass ratios (not only $m_u/m_c$, $m_c/m_t$, $m_d/m_s$ and
$m_s/m_b$, but also $m_u/m_d$) and Kobayashi-Maskawa matrix elements.
\end{quotation}

\newpage
Recent observation of top quark mass by the CDF Collaboration [1] has brought
a realistic study of quark mass matrix model within our reach more and more.
Now, of the ten independent observable quantities in three-quark-family scheme,
 we have already possessed experimental knowledge of nine quantities,
i.e. except for one parameter (a $CP$-violation phase parameter).
One of the criteria of model-building is how we can describe those observable
quantities with a few parameters as possible.
{}From the phenomenological point of view, for example, the Rosner-Worah
model [2] (and its democratic-type version [3]), which has six adjustable
parameters, can provide satisfactory predictions for six quark masses and
four independent observable quantities of Kobayashi-Maskawa (KM) [4] matrix.

Recently, one of the authors (Y.K.) has proposed a lepton and quark mass
matrix model [5]:
$$ M_f = m_0^f \, G O_f G \ ,  \eqno(1) $$
$$ G = {\rm diag}(g_1, g_2, g_3) \ , \eqno(2) $$
$$ O_f = {\bf 1}+ 3a_f X(\phi_f) \ , \eqno(3) $$
$$
{\bf 1}= \left(
\begin{array}{ccc}
1 & 0 & 0 \\
0 & 1 & 0 \\
0 & 0 & 1
\end{array} \right) \ , \qquad
X(\phi)=\frac{1}{3} \, \left(
\begin{array}{ccc}
1 & e^{i\phi} & 1 \\
e^{-i\phi} & 1 & 1 \\
1 & 1 & 1
\end{array} \right) \ , \eqno(4)
$$
where $f=\nu, e, u$, and $d$ are indices for neutrinos, charged leptons,
up- and down-quarks, respectively.
In the charged lepton mass matrix $M_e$, the parameter $a_e$ is chosen
as $a_e=0$, so that the parameters $m_0^e g_i^2$ are fixed by charged lepton
masses $m_i^e$ ($m_1^e=m_e$, $m_2^e=m_\mu$, $m_3^e=m_\tau$) as
$m_0^e g_i^2=m_i^e$.
Since the phase parameters $\phi_q$ are fixed at $\phi_u=0$ and $\phi_d=\pi/2$
and the parameters $m_0^u$ and $m_0^d$ are fixed as $m_0^u = m_0^d$,
the model includes only two adjustable parameters ($a_u$ and $a_d$) and
provides reasonable values of quark mass ratios (not absolute values) and
KM matrix parameters.

In spite of such phenomenological success, the following points in Ref. [5]
are still unsatisfactory to us: (i) $X(\phi_d)$ is not a rank one matrix
so that we must propose a complicated mechanism to explain the origin of
this term.  (ii) There are no reasonable explanation for nonzero phase terms
which exist only in (1, 2) and (2, 1) matrix elements of $X(\phi_d)$.  This
ansatz is contrary to the philosophy of ``democratic".  (iii) Even though
we accept that $O_d = {\bf 1} + 3 a_d X(\pi/2)$ is simple, the inverse matrix
$O_d^{-1}$ is not simple and it is difficult to account for
$ M_f = m_0^f \, G O_f G $ in a seesaw type model $M_f \simeq m M_F^{-1} m$
with $m \propto G$ and $ M_F \propto O_f^{-1}$.

In this paper we search for a possible form of the matrices $O_f$ in (1)
with the following conditions:  (a) The matrix form of each term in $O_f$
is as simple as possible.  (b) The number of hierarchically different terms
is as few as possible.  In a Higgs mechanism model, the latter condition
means Higgs fields are as few as possible.

The simplest form of $O_f$ is a unit matrix, but it leads to $M_u = M_d = M_e$
and fails to give the good predictions for quark mass spectrum and KM matrix.
The next simple form of $O_f$ is a unit matrix plus a rank one matrix, which
agrees with the condition (b).  Therefore it is meaningful to study quark
masses and mixing in the case that $O_q (q=u, d)$ is given by a unit matrix
plus a rank one matrix with a complex coefficient.

In the present paper, we propose the following quark mass matrix:
$$ M_f=(m_0^f/m_0^e)M_e^{1/2} O_f M_e^{1/2} \ ,  \eqno(5) $$
$$ M_e^{1/2}={\rm diag}(\sqrt{m_e},\sqrt{m_\mu},\sqrt{m_\tau}) \ , \eqno(6) $$
$$
O_f={\bf 1}+3a_f e^{i\alpha_f} X \ , \eqno(7)
$$
where $X \equiv X(0)$ is a rank one matrix.
The inverse matrices of $O_f$ are also simple
$$
O_f^{-1}={\bf 1}+3b_f e^{i\beta_f} X \ , \eqno(8)
$$
where
$$
b_f e^{i\beta_f} =- a_f e^{i\alpha_f}/(1+3a_f e^{i\alpha_f}) \ . \eqno(9)
$$
In a seesaw type model with heavy fermions, the inverse matrices $O_f^{-1}$
become more fundamental quantity than $O_f$ and the parameters $b_f$ and
$\beta_f$ are more important than $a_f$ and $\alpha_f$.

In the model of Ref. [5], there are only two adjustable parameters ($a_u$,
$a_d$) because of the ansatz $\phi_d=\pi/2$, but it is difficult to give a
reasonable explanation for the mass matrices in model building.
In the present model, three parameters ($a_u$, $a_d$, $\alpha_d$) are
necessary but a matrix form of $O_f^{-1}$ as well as $O_f$ are simple. This
makes model building easy.

If one feel the ``democratic" type matrix form $X$
in the present model (7) somewhat mysterious, one may alternatively
consider a diagonal matrix form diag$(0, 0, 1)$
by taking a suitable transformation of family basis,
because the unit matrix term {\bf 1} in (7) is unchanged
under this transformation.

The reason that we consider a democratic matrix from in $O_f$ is
motivated by only a phenomenological reason suggested in Ref. [6],
i.e., by the fact that for up-quark mass matrix with $\alpha_u=0$,
we can obtain the successful mass relation [6]
$${m_u}/{m_c} \simeq {3m_e}/{4m_\mu} \ , \eqno(10)$$
for a small value of $\varepsilon_u\equiv 1/a_u$.
Substituting the quark mass values which is given in eq. (12), the left
hand side of eq. (10) is $4.0 \times 10^{-3}$, while the right hand side
of eq. (10) is $3.6 \times 10^{-3}$.
Note that the ratio $m_u/m_c$ is insensitive to the parameter $a_u$.
The parameter $\varepsilon_u\equiv 1/a_u$ is determined by the mass ratio
$${m_c}/{m_t} \simeq 2(m_\mu/m_\tau)\varepsilon_u \ . \eqno(11)$$

The quark mass values [7] at an electroweak symmetry breaking energy scale
$\mu=\Lambda_W\equiv \langle\phi^0\rangle=
(\sqrt{2}G_F)^{-1/2}/\sqrt{2}=174$ GeV are
$$
\begin{array}[t]{lll}
m_u=0.0024\pm 0.0005\  {\rm GeV}  \ , & m_c=0.605\pm 0.009 \  {\rm GeV}  \ ,
& m_t=174\pm 10 ^{+13}_{-12} \ {\rm GeV}  \ , \\
 m_d=0.0042\pm 0.0005 \  {\rm GeV} \ , & m_s=0.0851\pm 0.014 \  {\rm GeV}  \ ,
& m_b=2.87\pm 0.03 \  {\rm GeV}  \ , \\
\end{array}
 \eqno(12)
$$
where $\langle\phi^0\rangle$ is a vacuum expectation value of a Higgs scalar
field $\phi^0$ in the standard model and we have used
$\Lambda_{\overline{MS}}^{(4)}=0.26$ GeV.

Differently from the model given in Ref. [5], down-quark mass matrix
$M_d$ with $\alpha_d\neq 0$ in the present model is not Hermitian.
We will demonstrate that the present model with the form (7) also
can provide reasonable predictions of quark mass ratios and KM matrix
by adjusting our parameters $a_u$, $a_d$ and $\alpha_d$ (i.e.,
$b_u$, $b_d$ and $\beta_d$).

In the present model, a case $a_d\simeq -1/2$ can provide
phenomenologically interesting predictions as seen below.
For small values of $|\alpha_d|$ and $\varepsilon_d\equiv -(2+a_d^{-1})$,
we obtain the down-quark mass ratios
$$
\frac{m_s}{m_b} \simeq \frac{1}{2} \kappa\left(1-
48\sqrt{\frac{2m_e m_\mu}{3m_\tau^2}}\right) \ , \eqno(13)
$$
$$
\frac{m_d}{m_s} \simeq \frac{16}{\kappa^2}\frac{m_e m_\mu}{m_\tau^2}
\left(1+96\sqrt{\frac{2m_e m_\mu}{3m_\tau^2}}\right) \ , \eqno(14)
$$
where
$$
\kappa = \sqrt{\sin^2\frac{\alpha_d}{2} +
\left(\frac{\varepsilon_d}{4}\right)^2} \ . \eqno(15)
$$
We also obtain
$$
\frac{m_d m_s}{m_b^2}\simeq 4 \frac{m_e m_\mu}{m_\tau^2} \ , \eqno(16)
$$
as a relation which is insensitive to the small parameters $|\alpha_d|$
and $\varepsilon_d$.
The left hand side of eq. (16) is $4.3 \times 10^{-5}$, while the right hand
side of eq. (16) is $6.8 \times 10^{-5}$ with the quark mass values (12).

Furthermore, we can obtain ratios of up-quarks to down-quarks,
for example,
$$
{m_u}/{m_d} \simeq 6 \kappa \sim 12 {m_s}/{m_b} \ . \eqno(17)
$$
Suitable choice of small values of $\varepsilon_d$ and
$\alpha_d$ ensures $m_u/m_d\sim O(1)$ in spite of $m_t\gg m_b$.
{}From (9), a small value $|\varepsilon_u|=1/|a_u|\simeq 0$ means
$b_u\simeq -1/3$, while a small value
$|\varepsilon_d|=|2+a_d^{-1}|\simeq 0$ means $b_d\simeq -1$.
It is noted that, in spite of the large ratio of $m_t/m_b$,
the ratio of $b_d/b_u$ is not so large, i.e., $b_d/b_u\simeq 3$.


Then, let us discuss the KM matrix elements $V_{ij}$.
The KM matrix $V$ is given by
$$
V= U_L^u P U_L^{d\dagger} \ , \eqno(18)
$$
where $U_L^u$ and $U_L^d$ are defined by
$$
U_L^u M_u M_u^\dagger U_L^{u\dagger}= {\rm diag}(m_u^2, m_c^2, m_t^2) \ , \quad
U_L^d M_d M_d^\dagger U_L^{d\dagger}= {\rm diag}(m_d^2, m_s^2, m_b^2)
\ , \eqno(19)
$$
respectively, and $P$ is a phase matrix.
Here, we have considered that the quark basis for the mass matrix (5) can,
in general, deviate from the quark basis of weak interactions
by some phase rotations,
The simplest case $P=$diag$(1,1,1)$ cannot provide reasonable predictions
of $|V_{ij}|$.
When we take
$$P={\rm diag}(1,1,-1) \ , \eqno(20) $$
we can obtain reasonable predictions for
both quark mass ratios and KM matrix elements,
although it is an open question why such a phase inversion is caused on
the third family quark.
The predictions of $|V_{ij}|$ are sensitive to every value of
$\varepsilon_u$, $\varepsilon_d$ and $\alpha_d$, so that it is not
adequate to express $|V_{ij}|$ as simple approximate relations
such as those in (10)--(11) and (13)--(17).
Therefore, we will show only numerical results for $|V_{ij}|$.
For example, by taking
$a_u=28.65$, $a_d=-0.4682$, $\alpha_u=0$ and $\alpha_d=7.96^\circ$
($b_u=-0.3295$, $b_d=-1.072$, $\beta_u=0$ and
$\beta_d=18.5^\circ$), which are chosen by fitting the quark mass ratios,
we obtain the following predictions of quark masses, KM matrix elements
$|V_{ij}|$ and the rephasing-invariant quantity $J$ [8]:
$$
\begin{array}{lll}
m_u=0.00228\ {\rm GeV} \ , & m_c=0.591 \ {\rm GeV} \ , &
m_t=170 \ {\rm GeV} \ , \\
m_d=0.00429\ {\rm GeV} \ , & m_s=0.0875 \ {\rm GeV} \ , &
m_b=3.02 \ {\rm GeV} \ ,  \\
\end{array}   \eqno(21)
$$
$$
\begin{array}{lll}
|V_{us}|=0.223 \ , & |V_{cb}|=0.0542 \ , & |V_{ub}|=0.00309 \ , \\
|V_{td}|=0.0146 \ , & |V_{ub}/V_{cb}|=0.0570 \ , & J=2.30 \times 10^{-5} \ . \\
\end{array}   \eqno(22)
$$
The prediction $|V_{cb}|=0.0542$ in (22) is somewhat large in comparison
with the experimental value $|V_{cb}|=0.040\pm 0.005$ [9].
If we use $P=(1,1,-e^{i\delta})$ with a small phase value $\delta$
instead of $P=(1,1,-1)$,
we can obtain more excellent predictions without changing
predictions of quark masses in (21):
for example, when we take $\delta=-4.4^\circ$, we obtain
$$
\begin{array}[t]{lll}
|V_{us}|=0.223 \ , & |V_{cb}|=0.0400  \ , & |V_{ub}|=0.00274 \ ,  \\
|V_{td}|=0.0111 \ , & |V_{ub}/V_{cb}|=0.0686 \ , & J=1.55 \times 10^{-5} \ . \\
\end{array}
  \eqno(23)
$$

In the numerical predictions of quark masses (21),
we have used a common enhancement factor
of quark masses to lepton masses, $m_0^u/m_0^e=m_0^d/m_0^e=3$,
in order to compare with quark mass values at $\mu=\Lambda_W$ (12).
It is an open question why we can set the factor $m_0^q/m_0^e$ as just three.
Although we are happy if we can explain such the factor $m_0^q/m_0^e=3$
by evolving quark and lepton masses from $\mu=\Lambda_X$
to $\mu=\Lambda_W$,
unfortunately, it is not likely to derive such a large factor $\sim 3$
from the conventional renormalization calculation.

At present, we have no theory to determine the parameters
$a_f$ and $\alpha_f$.
For charged leptons, we must take $a_e=0$.
For quarks, we have chosen $a_q$ from the phenomenological parameter fitting.
However, in the present stage, we do not provide any unified understanding for
$a_f$ and $\alpha_f$, i.e., they are nothing more than
phenomenological parameters.


In conclusion,
quark mass ratios and KM matrix elements can be fitted only by three
parameters $a_u$, $a_d$ and $\alpha_d$ ($b_u$, $b_d$ and $\beta_d$) fairly
well. If we take a seesaw-type model, we must consider that the parameters
$b_f$ and $\beta_f$ in $O_f^{-1}$ are more fundamental ones rather than
$a_f$ and $\alpha_f$ in $O_f$. Then
it is worth while that we can obtain a large ratio of $m_t/m_b$
together with a reasonable ratio $m_u/m_d$ without taking  so
hierarchically different values of $b_u$ and $b_d$, i.e.,
with taking $b_u\simeq -1/3$ and $b_d \simeq -1$, in contrast to $a_u
\simeq 30$ and $a_d \simeq -1/2$ in the case of $GO_fG$ picture.
\vglue.3in

\centerline{\bf Acknowledgments}
The authors are grateful to Professor M.~Tanimoto for the simulating comments.
This work was supported by the Grant-in-Aid for Scientific Research,
Ministry of Education, Science and Culture, Japan (No.06640407).

\vglue.3in
\newcounter{0000}
\centerline{\bf References and Footnote}
\begin{list}
{[~\arabic{0000}~]}{\usecounter{0000}
\labelwidth=0.8cm\labelsep=.1cm\setlength{\leftmargin=0.7cm}
{\rightmargin=.2cm}}
\item  CDF Collaboration, F.~Abe {\it et al}., Phys.~Rev.~Lett. {\bf 73},
225 (1994); Phys. Rev. {\bf D50}, 2966 (1994).
\item J. L. Rosner and M. Worah, Phys.~Rev. {\bf D46}, 1131 (1992).
\item Y.~Koide, in {\it International Symposium on Extended Objects and
Bound Systems}, Proceedings, Karuizawa, Japan, 1992,
edited by O.~Hara, S.~Ishida, and S.~Naka
(World Scientific, Singapore, 1992);
K.~Matumoto, Prog.~Theor.~Phys. {\bf 89}, 269 (1993);
Y. Koide and H. Fusaoka, Phys.~Rev. {\bf D48}, 432 (1993).
\item M.~Kobayashi and T.~Maskawa, Prog.~Theor.~Phys. {\bf 49}, 652 (1973).
\item Y.~Koide, Phys.~Rev. {\bf D49}, 2638 (1994).
\item Y.~Koide, Mod.~Phys.~Lett. {\bf A8}, 2071 (1993).
\item For light quark masses $m_q(\mu)$, we have used the values at $\mu=1$
GeV, $m_u=5.6\pm 1.1$ MeV, $m_d=9.9\pm 1.1$ MeV and $m_s=199\pm 33$ MeV:
C.~A.Dominquez and E.~de Rafael, Annals of Physics {\bf 174}, 372 (1987).
However, the absolute values of light quark masses should be taken
solidly because they depend on models.
For $m_c(\mu)$ and $m_b(\mu)$, we have used the value $m_c(m_c)=1.26\pm
0.02$ GeV by Narison, and the value $m_b(m_b)=4.72\pm 0.05$ GeV
by Dominquez--Paver: S.~Narison, Phys.~Lett. {\bf B216}, 191 (1989);
C.~A.~Dominquez and N.~Paver, Phys.~Lett. {\bf B293}, 197 (1992).
For top quark mass, we have used $m_t(m_t)=174\pm 10 ^{+13}_{-12}$ GeV
from the CDF experiment [1].
\item C.~Jarlskog, Phys.~Rev.~Lett. {\bf 55}, 1839 (1985);
O.~W.~Greenberg, Phys.~Rev. {\bf D32}, 1841 (1985);
I.~Dunietz, O.~W.~Greenberg, and D.-d.~Wu, Phys.~Rev.~Lett. {\bf 55},
2935 (1985);
C.~Hamzaoui and A.~Barroso, Phys.~Lett. {\bf 154B}, 202 (1985);
D.-d.~Wu, Phys.~Rev. {\bf D33}, 860 (1986).
\item Particle Data Group, L.~Montanet et al., Phys. Rev. {\bf D50},
1173 (1994).
\end{list}

\end{document}